\documentclass[sigconf,nonacm]{acmart}
\AtBeginDocument{%
  }

\copyrightyear{2025}
\acmYear{2025}
\setcopyright{rightsretained}
\acmConference[CF Companion '25]{22nd ACM International Conference on Computing Frontiers}{May 28--30, 2025}{Cagliari, Italy}
\acmBooktitle{22nd ACM International Conference on Computing Frontiers (CF Companion '25), May 28--30, 2025, Cagliari,Italy}\acmDOI{10.1145/3706594.3726978}
\acmISBN{979-8-4007-1393-4/2025/05}

\usepackage[acronym]{glossaries}
\usepackage{multirow}

\newacronym{HPC}{HPC}{high-performance computing}
\newacronym{ML}{ML}{machine learning}
\newacronym{FPU}{FPU}{floating-point unit}
\newacronym{VRF}{VRF}{vector register file}
\newacronym{MRF}{MRF}{matrix register file}
\newacronym{RF}{RF}{register file}
\newacronym{PPA}{PPA}{power, performance, and area}
\newacronym{MAC}{MAC}{multiply–accumulate}
\newacronym{LSU}{LSU}{Load-Store Unit}
\newacronym{AU}{AU}{arithmetic unit}
\newacronym{SA}{SA}{systolic array}
\newacronym{matmul}{MatMul}{matrix multiplication}
\newacronym{rvv}{RVV}{RISC-V Vector}
\newacronym{XIF}{XIF}{OpenHW Group CORE-V-X interface}

\newcommand{\QTimeQCYCspf}  {\ensuremath{17676}}
\newcommand{\QTimeQCYCspi}  {\ensuremath{17676}}
\newcommand{\QTimeQCYChpi}  {\ensuremath{9484}}
\newcommand{\QTimeQCYCbpi}  {\ensuremath{5388}}
\newcommand{\QTimeQPIspf}   {\ensuremath{98.5\%}}
\newcommand{\QTimeQPIspi}   {\ensuremath{98.5\%}}
\newcommand{\QTimeQPIhpi}   {\ensuremath{97.2\%}}
\newcommand{\QTimeQPIbpi}   {\ensuremath{93.2\%}}
\newcommand{\QTimeQFUspf}   {\ensuremath{92.7\%}}
\newcommand{\QTimeQFUspi}   {\ensuremath{92.7\%}}
\newcommand{\QTimeQFUhpi}   {\ensuremath{86.4\%}}
\newcommand{\QTimeQFUbpi}   {\ensuremath{76.0\%}}

\newcommand{\QTimeKCYCspf}  {\ensuremath{4120}}
\newcommand{\QTimeKCYCspi}  {\ensuremath{4120}}
\newcommand{\QTimeKCYChpi}  {\ensuremath{2072}}
\newcommand{\QTimeKCYCbpi}  {\ensuremath{1048}}
\newcommand{\QTimeKPIspf}   {\ensuremath{99.8\%}}
\newcommand{\QTimeKPIspi}   {\ensuremath{99.8\%}}
\newcommand{\QTimeKPIhpi}   {\ensuremath{99.2\%}}
\newcommand{\QTimeKPIbpi}   {\ensuremath{98.1\%}}
\newcommand{\QTimeKFUspf}   {\ensuremath{99.4\%}}
\newcommand{\QTimeKFUspi}   {\ensuremath{99.4\%}}
\newcommand{\QTimeKFUhpi}   {\ensuremath{98.8\%}}
\newcommand{\QTimeKFUbpi}   {\ensuremath{97.7\%}}

\newcommand{\QTimeSCYCspf}  {\ensuremath{5398}}
\newcommand{\QTimeSCYCspi}  {\ensuremath{5398}}
\newcommand{\QTimeSCYChpi}  {\ensuremath{3340}}
\newcommand{\QTimeSCYCbpi}  {\ensuremath{2316}}
\newcommand{\QTimeSPIspf}   {\ensuremath{94.8\%}}
\newcommand{\QTimeSPIspi}   {\ensuremath{94.8\%}}
\newcommand{\QTimeSPIhpi}   {\ensuremath{92.0\%}}
\newcommand{\QTimeSPIbpi}   {\ensuremath{88.4\%}}
\newcommand{\QTimeSFUspf}   {\ensuremath{75.9\%}}
\newcommand{\QTimeSFUspi}   {\ensuremath{75.9\%}}
\newcommand{\QTimeSFUhpi}   {\ensuremath{61.3\%}}
\newcommand{\QTimeSFUbpi}   {\ensuremath{44.2\%}}

\newcommand{\QTOTAreaMM} {\ensuremath{0.65}}
\newcommand{\QTOTArea}   {\ensuremath{652788}}
\newcommand{\QCtrlArea}  {\ensuremath{20670}}
\newcommand{\QPUArea}    {\ensuremath{235}}
\newcommand{\QLSUArea}   {\ensuremath{17231}}
\newcommand{\QRFArea}    {\ensuremath{74510}}
\newcommand{\QSAArea}    {\ensuremath{540142}}

\newcommand{\QSAAreaComb}    {\ensuremath{462861}}
\newcommand{\QSAAreaSeq}     {\ensuremath{77281}}

\newcommand{\QCtrlAreaP}    {\ensuremath{3.1\%}}
\newcommand{\QRFAreaP}      {\ensuremath{11.4\%}}
\newcommand{\QPUAreaP}      {\ensuremath{0.1\%}}
\newcommand{\QLSUAreaP}     {\ensuremath{2.6\%}}
\newcommand{\QSAAreaP}      {\ensuremath{82.8\%}}
\newcommand{\QSAAreaSeqP}   {\ensuremath{11.8\%}}
\newcommand{\QSAAreaCombP}  {\ensuremath{71.0\%}}

\newcommand{\AQsameFPU}          {\ensuremath{33\%}}

\newcommand{\AQsameFPUAEffP}     {\ensuremath{58\%}}

\newcommand{\EQsameFPU}          {\ensuremath{6\%}}
\newcommand{\TQsameFPU}          {\ensuremath{0.1\%}}

\newcommand{\AQsameBWAEffP}     {\ensuremath{62\%}}
\newcommand{\EQsameBW}          {\ensuremath{15\%}}
\newcommand{\TQsameBW}          {\ensuremath{3.87}}

\newcommand{\AQMXAEff}      {\ensuremath{1.77}}
\newcommand{\AQMXAEffP}     {\ensuremath{77\%}}
\newcommand{\EQMX}          {\ensuremath{13\%}}
\newcommand{\TQMX}          {\ensuremath{3.86}}

\begin{document}

\title{Quadrilatero: A RISC-V programmable matrix coprocessor for low-power edge applications}

\author{Danilo Cammarata}
\orcid{0009-0003-8147-758X}
\affiliation{
  \institution{ETH Zürich}
  \city{Zürich}
  \country{Switzerland}
}
\email{dcammarata@iis.ee.ethz.ch}

\author{Matteo Perotti}
\orcid{0000-0003-2413-8592}
\affiliation{
  \institution{ETH Zürich}
  \city{Zürich}
  \country{Switzerland}
}
\email{mperotti@iis.ee.ethz.ch}

\author{Marco Bertuletti}
\orcid{0000-0001-7576-0803}
\affiliation{
  \institution{ETH Zürich}
  \city{Zürich}
  \country{Switzerland}
}
\email{mbertuletti@iis.ee.ethz.ch}

\author{Angelo Garofalo}
\orcid{0000-0002-7495-6895}
\affiliation{%
  \institution{ETH Zürich}
  \city{Zürich}
  \country{Switzerland}
}
\affiliation{%
  \institution{Università di Bologna}
  \city{Bologna}
  \country{Italy}
}
\email{agarofalo@iis.ee.ethz.ch}

\author{Pasquale Davide Schiavone}
\orcid{0000-0003-2931-0435}
\author{David Atienza}
\orcid{0000-0001-9536-4947}
\affiliation{%
  \institution{EPFL}
  \city{Lausanne}
  \country{Switzerland}
}
\email{davide.schiavone@epfl.ch}
\email{david.atienza@epfl.ch}

\author{Luca Benini}
\orcid{0000-0001-8068-3806}
\affiliation{%
  \institution{ETH Zürich}
  \city{Zürich}
  \country{Switzerland}
}
\affiliation{%
  \institution{Università di Bologna}
  \city{Bologna}
  \country{Italy}
}
\email{lbenini@iis.ee.ethz.ch}

\begin{abstract}
  The rapid growth of AI-based Internet-of-Things applications increased the demand for high-performance edge processing engines on a low-power budget and tight area constraints. As a consequence, vector processor architectures, traditionally designed for \acrfull{HPC}, made their way into edge devices, promising high utilization of \acrlong{FPU}s (\acrshort{FPU}s) and low power consumption. 
  However, vector processors can only exploit a single dimension of parallelism, leading to expensive accesses to the \acrfull{VRF} when performing matrix computations, which are pervasive in AI workloads.
  To overcome these limitations while guaranteeing programmability, many researchers and companies are developing dedicated instructions for a more efficient \acrfull{matmul} execution.
  In this context, we propose Quadrilatero, an open-source RISC-V programmable systolic array coprocessor for low-power edge applications that implements a streamlined matrix ISA extension.
  We evaluate the post-synthesis \acrfull{PPA} metrics of Quadrilatero in a mature 65-nm technology node, showing that it requires only {\QTOTAreaMM} $mm^2$ and that it can reach up to {\QTimeKFUspf} of \acrshort{FPU} utilization. 
  Compared to a state-of-the-art open-source RISC-V vector processor and a hybrid vector-matrix processor optimized for embedded applications, Quadrilatero improves area efficiency and energy efficiency by up to {\AQMXAEffP} and {\EQsameBW}, respectively.
\end{abstract}





\maketitle
\section{Introduction}
In recent years, \acrfull{ML} models have become increasingly widespread in edge computing and AI applications \cite{TinyML}.
These models rely on computationally intensive algorithms that process parallel workloads using vector and matrix operators, with \acrshort{matmul} being the dominant one.

Deployed initially in supercomputers, vector processor architectures have recently proven to be a valid and efficient programmable solution to perform these tasks even in the edge domain \cite{spatz,SPEED} and to adapt to novel algorithms, mitigating the risk of becoming outdated.
Despite providing high utilization and efficiency when computing matrix workloads, vector instructions can only exploit parallelism in one dimension at a time. Hence, they achieve suboptimal efficiency on multi-dimensional data structures (e.g., matrices) when none of the dimensions is large enough to amortize instruction fetch and setup. Furthermore, they impose a 
high bandwidth requirement between the \acrshort{VRF} and the \acrshort{FPU}s when scaling up.

To overcome these limitations, multiple ISA vendors have proposed matrix ISA extensions, such as Arm SME, Intel AMX, and IBM MMA, to improve the execution of matrix workloads.
Even if RISC-V has not ratified a matrix ISA extension yet, the standardization effort led by a working task group is ongoing, and several companies have already released ISA extension proposals \cite{STC,THEAD,SpaceMIT}.

State-of-the-art accelerators for high arithmetic-intensity computations based on a systolic array \cite{Kung}, such as Gemmini \cite{Gemmini}, CONNA \cite{CONNA}, and OpenGeMM \cite{OpenGeMM}, have an area footprint and power cost that exceed the budget of low-power edge platforms (respectively, 2.4 $\mathrm{mm}^2$, 2.36 $\mathrm{mm}^2$ and 2.6 $\mathrm{mm}^2$ in a 65-nm node).\\
In this work, we focus on the integration of a systolic array coprocessor for a RISC-V processor in a low-power microcontroller class configuration for low-power edge matrix-intensive applications. Our contributions are:  \begin{itemize}
\item The ISA and architectural specification of Quadrilatero\footnote{Quadrilatero GitHub: https://github.com/pulp-platform/quadrilatero/} and its design and integration as a coprocessor of an RVI32 core.
\item The post-synthesis \acrfull{PPA} analysis of Quadrilatero in a 65-nm low-power technology node. Quadrilatero requires {\QTOTAreaMM} $mm^2$, consumes 34 mW at 100 MHz when executing a \acrshort{matmul} between 64$\times$64 matrices and reaches up to {\QTimeKFUspf} \acrshort{FPU} utilization.
\item A comparison of Quadrilatero with Spatz \cite{spatz}, a RISC-V vector processor optimized for low-power edge applications. Compared to a Spatz with the same register file bandwidth, Quadrilatero improves the area efficiency by {\AQsameBWAEffP}, the execution time by {\TQsameBW}$\times$ and reduces the energy consumption by {\EQsameBW}. Compared to a Spatz with the same number of \acrshort{FPU}s, it reaches comparable execution time, improves the area efficiency by {\AQsameFPUAEffP}, and the energy consumption by {\EQsameFPU}.
\item A comparison of Quadrilatero with Spatz MX \cite{spatzMX}, a hybrid vector-matrix processor for embedded applications. 
Quadrilatero improves the area efficiency by {\AQMXAEffP}, the execution time by {\TQMX}$\times$, and reduces the energy consumption by {\EQMX}.
\end{itemize}

\section{Background}
\begin{figure}[h]
    \centering
      \raggedright
      \footnotesize
      \ttfamily
       \begin{tabbing}
        \hspace{1em} \= \hspace{1em} \= \hspace{1em} \= \hspace{1em} \= \kill
        1.\>for (int m = 0; m < M; m += 8) \{ \\
        2.\>\> for (int n = 0; n < N; n += 8) \{ \\
        3.\>\>\> mz m4; mz m5; mz m6; mz m7; \\
        4.\>\>\> for (int k = 0; k < K; k += 4) \{ \\
        5.\>\>\>\> mld.w     m0, \&mtxA[m*M+k]; \\
        6.\>\>\>\> mld.w     m1, \&mtxB[n*N+k];  // transposed in memory.\\
        7.\>\>\>\> mmac      m4, m0, m1; \\
        8.\>\>\>\> mld.w       m2, \&mtxA[(m+4)*M+k];\\
        9.\>\>\>\> mmac      m6, m2, m1; \\
        10.\>\>\>\> mld.w     m3, \&mtxB[(n+4)*N+k]; // transposed in memory.\\
        11.\>\>\>\> mmac      m5, m0, m3; \\
        12.\>\>\>\> mmac      m7, m2, m3; \\
        13.\>\>\> \} \\
        14.\>\>\> mst.w     m4, \&mtxC[m*M+n]; mst.w     m5, \&mtxC[m*M+n+4];\\
        15.\>\>\> mst.w     m6, \&mtxC[(m+4)*M+n]; mst.w     m7, \&mtxC[(m+4)*M+n+4];\\
        16.\>\>\} \\
        17.\>\}
    \end{tabbing}
    \includegraphics[width=0.45\textwidth]{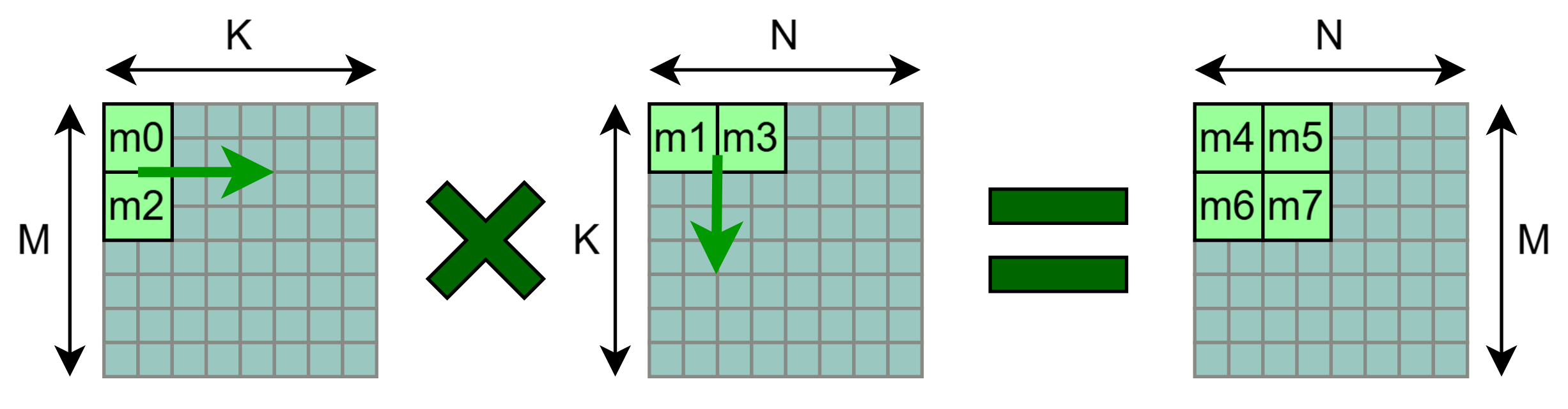}
    \caption{Pseudocode of a 8x8-based \acrshort{matmul} with matrix instructions and its graphical representation.}
    \label{fig:pseudocode}
\end{figure}
The key difference between a vector ISA and a matrix ISA in micro-architectural terms lies in the bandwidth requirements between the \acrfull{RF} and \acrshort{FPU}s. The \acrfull{rvv} instruction $vfmacc.vv$ performs $VLEN/SEW$ \acrfull{MAC} operations\footnote{VLEN: vector register length [bit]. SEW: element width [bit].} by moving $4 \times VLEN/SEW$ elements between \acrshort{VRF} and \acrshort{FPU}s, while a matrix  \acrshort{MAC} instruction can relax this strong requirement on \acrshort{VRF} bandwidth.
The matrix ISA extension we implement in Quadrilatero defines eight matrix registers (\textit{m0},...,\textit{m7}), each made of ${RLEN}/{32}$ rows with $RLEN$ bits per row. The core instructions implemented in Quadrilatero are shown in Figure \ref{fig:pseudocode}: $mz$ (line 3) resets a matrix register; $mld.w$ (line 5) loads 32-bit values into a matrix register; $mst.w$ (line 14) stores the 32-bit values from a matrix register. In addition to these initialization and data-transfer instructions, the extension defines $mmac$ (line 7), namely \acrshort{MAC} instructions that depend on the data types and require three matrix registers, one of which holds transposed values. The $mmac$ encodes $( {RLEN/32})^2 \times {RLEN}/{SEW}$ \acrshort{MAC} operations and moves $ 4 \times {RLEN}/{32} \times {RLEN}/{SEW}$ elements from/to the \acrshort{RF}, thus reducing the number of \acrshort{RF} accesses by ${RLEN}/{32}$ compared to $vfmacc.vv$. Moreover, compared to an \acrshort{rvv} instance with $DLEN$ bits as \acrshort{RF}-to-\acrshort{FPU}s bandwidth, which can reach up to $DLEN/SEW$ \acrshort{MAC}s/cycle, our matrix ISA can increase the \acrshort{MAC}s/cycle by ${RLEN}/{32}$ given the same \acrshort{RF}-to-\acrshort{FPU}s bandwidth.
Specifically, we configure Quadrilatero with $RLEN=128$ bits, achieving up to 16 MACs/cycle.
\section{Architecture}
\begin{figure}[h]
    \centering
    \includegraphics[width=0.28\textwidth]{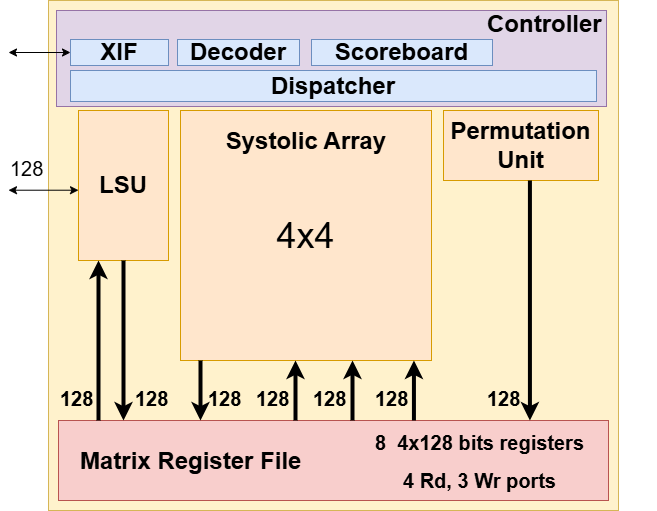}
    \caption{Quadrilatero Architecture with RLEN = 128.}
    \label{fig:architecture}
\end{figure}
\begin{figure*}[t]
    \centering
    \includegraphics[width=0.9\textwidth]{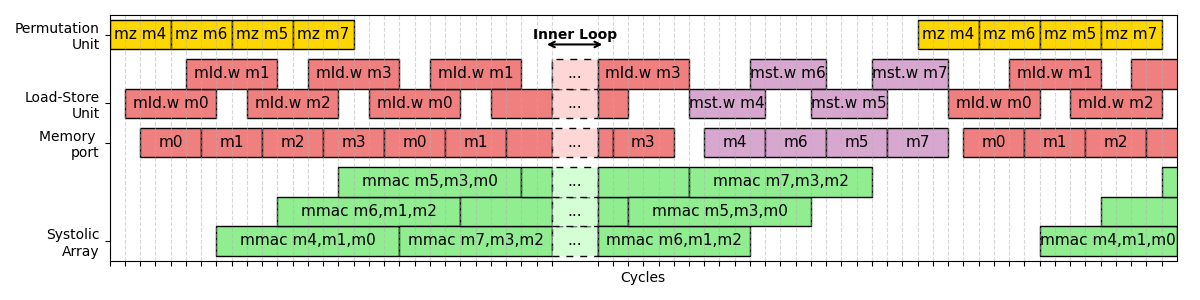}
    \vspace{-0.5cm}
    \caption{Gantt Chart of the intermediate loop of the \acrshort{matmul} kernel executed by Quadrilatero. The inner loop is executed without stalls, while two consecutive intermediate loop iterations have only three cycles of losses on the memory port.}
    \label{fig:gantt}
\end{figure*}
Quadrilatero's architecture is shown in Figure \ref{fig:architecture}. It interfaces with a 32-bit scalar RISC-V core through the \gls{XIF}. The scalar core offloads the matrix instructions to Quadrilatero (along with the scalar operands) and waits for their completion before committing them. Consequently, Quadrilatero has its own decoder and can dispatch the instructions to three different execution units: the Permutation Unit, which executes the $mz$ instruction to reset matrix registers; the \acrfull{LSU} for memory operations; and the \acrfull{SA} for \acrshort{MAC} operations. Quadrilatero also has a scoreboard to track all data dependencies and guarantee correct access to the \acrfull{MRF}.

Since Quadrilatero targets the low-power edge domain, we designed it to keep its area below 1 $mm^2$ in 65-nm technology while maximizing the number of \acrshort{MAC}s per cycle on 32-bit data. From an explorative synthesis, we know that these conditions are satisfied by $RLEN=128$ bits when each matrix register holds a 4x4 matrix.

Figure \ref{fig:gantt} shows the instruction scheduling in the execution units during a \acrshort{matmul}.
To prevent execution units from stalling, the \acrshort{MRF} has four dedicated read ports and three dedicated write ports (each 128-bit wide). Each matrix register is accessible row by row so that it can be read/written in four cycles.

To balance the execution time of memory and arithmetic instructions in the inner loop of the \acrshort{matmul} kernel shown in Figure \ref{fig:pseudocode}, we matched \acrshort{MRF} bandwidth, \acrshort{SA} throughput, and memory bandwidth.
Consequently, we designed the \acrshort{SA} as a 4x4 32-bit grid of  \acrshort{MAC} units. Each MAC unit can support integer SIMD operations on 32-bit accumulators with 8-bit, 16-bit, and 32-bit input operands and fp32 \acrshort{MAC} operations, operating in a single cycle. To maximize \acrshort{MAC} unit utilization, we implemented the \acrshort{SA} based on a weight-stationary flow inspired by the Weight-Load-Skip with Double-Buffering (WLS-DB) Register Aware Systolic Array \cite{RASA}. This flow comprises three independent stages, enabling the execution of up to three instructions in parallel. Thus, the \acrshort{SA} requires 12 cycles to complete the execution of a single $mmac$, but it can execute consecutive $mmac$ in four cycles on average, reaching 16 \acrshort{MAC}s/cycle at full capacity.
The \acrshort{LSU} has a 128-bit/cycle memory bandwidth and dedicated \acrshort{MRF} ports to enable simultaneous execution of two $mld$ or two $mst$. The \acrshort{MRF}-to-memory path is cut with buffers to decouple memory and \acrshort{MRF}. Data moved between \acrshort{VRF} and memory is always stored in the corresponding buffer before reaching its destination. To prevent data hazards, $mld$ and $mst$ cannot be executed in parallel. On average, memory operations take four cycles when no stalls occur. 
As shown in Figure \ref{fig:gantt}, these design choices lead to fully utilizing the \acrshort{SA} and the memory port when executing the inner loop of a \acrshort{matmul} kernel and to have only three cycles lost on the memory port for each intermediate loop iteration.

\section{Experimental results}
\begin{figure*}[t]
    \centering
    \begin{minipage}{0.4\textwidth}  
    \centering
        \includegraphics[width=0.4\textwidth]{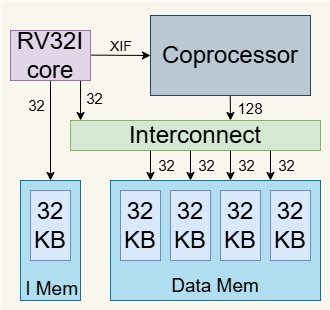}
        \label{fig:general_system}
    \hspace{0.5cm}  
        \includegraphics[width=0.4\textwidth]{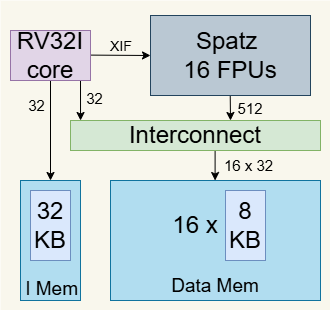}
        \label{fig:spatz16}
    \caption{On the left, the system where we integrate Quadrilatero, Spatz with 4 \acrshort{FPU}s and Spatz MX. On the right, the system where we integrate Spatz with 16 \acrshort{FPU}s.}
    \label{fig:main}
    \end{minipage}
    \hfill
    \begin{minipage}{0.45\textwidth}  
        \includegraphics[width=\textwidth]{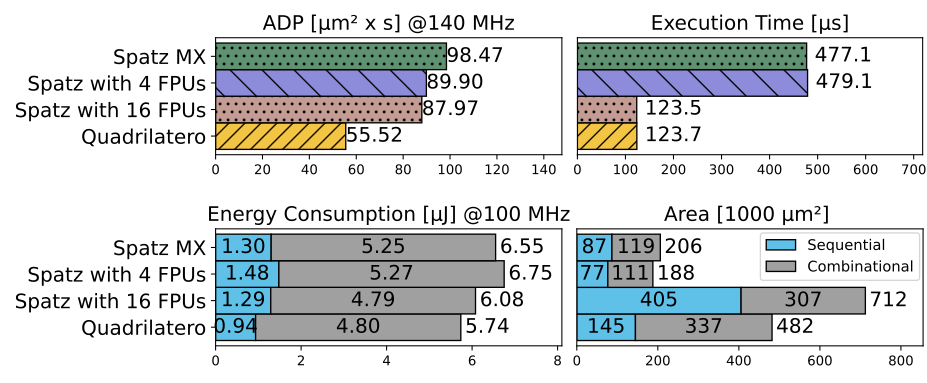}
        \caption{Experimental results on the comparison of the register file and \acrshort{FPU} of the different systems.}
        \label{fig:comparison}
    \end{minipage}
\end{figure*}

We couple Quadrilatero to the \texttt{RV32I} CV32E40PX scalar core (i.e., a CV32E40P core \cite{cv32e40p} extended with the \gls{XIF}) and integrate them in a multi-banked memory system with four 32-KiB interleaved data memory banks, as shown in Figure \ref{fig:main}.
We measure the cycle runtime of three different \acrshort{matmul} workloads that can fit into our memory system and fully exploit Quadrilatero's resources (\acrshort{MRF} and 16 \acrshort{MAC} units) for all the data types supported by Quadrilatero. 
Specifically, we select the largest problem size ($M\times K\times N$) with square matrices ($64\times64\times64$), with the highest \texttt{K} ($8\times 1024\times 8$ ), and with the lowest \texttt{K} ($64\times 16\times 64$). 
In Table \ref{tab:q_perf}, we report the number of cycles, the performance ideality (i.e., the ratio between the minimum theoretical number of cycles required by a workload\textemdash given a specific memory bandwidth and number of \acrshort{MAC} units\textemdash and the achieved number of cycles), and the \acrshort{FPU} utilization.
As shown in Figure \ref{fig:gantt}, the execution of $mld.w$ is balanced with the $mmac$ execution. Thus, Quadrilatero is limited by the $mst.w$, leading to an \acrshort{FPU} utilization lower than the maximum theoretical performance. 
This overhead is constant and depends on \texttt{K}: the higher the \texttt{K}, the lower the overhead. In particular, when \texttt{K}=1024, the \acrshort{FPU} utilization reaches \textbf{\QTimeKFUspf}. Narrower data types incur lower performance since Quadrilatero processes narrow data in SIMD fashion, effectively lowering the number of iterations over the \texttt{K} dimension.
\begin{table}[t]
    \centering
    \small
    \caption{Quadrilatero's performance with different \acrshort{matmul} workloads.}
          \begin{tabular}{lcccc}
            \toprule
            \multirow{2}{*}{Data types} & Matrix Sizes          & \multirow{2}{*}{Cycles} & Performance & \acrshort{FPU} \\
                                        & (M $\times$ K $\times$ N) &                         & Ideality    & Utilization    \\
            \midrule
            $\text{fp32} \to \text{fp32}$   & $64 \times 64 \times 64$ & {\QTimeQCYCspf} & {\QTimeQPIspf} & {\QTimeQFUspf}\\
            $\text{int32} \to \text{int32}$ & $64 \times 64 \times 64$ & {\QTimeQCYCspi} & {\QTimeQPIspi} & {\QTimeQFUspi}\\
            $\text{int16} \to \text{int32}$ & $64 \times 64 \times 64$ & {\QTimeQCYChpi} & {\QTimeQPIhpi} & {\QTimeQFUhpi}\\
            $\text{int8} \to \text{int32}$  & $64 \times 64 \times 64$ & {\QTimeQCYCbpi} & {\QTimeQPIbpi} & {\QTimeQFUbpi}\\
            \midrule
            $\text{fp32} \to \text{fp32}$   & $8 \times 1024 \times 8$ & {\QTimeKCYCspf} & {\QTimeKPIspf} & {\QTimeKFUspf}\\
            $\text{int32} \to \text{int32}$ & $8 \times 1024 \times 8$ & {\QTimeKCYCspi} & {\QTimeKPIspi} & {\QTimeKFUspi}\\
            $\text{int16} \to \text{int32}$ & $8 \times 1024 \times 8$ & {\QTimeKCYChpi} & {\QTimeKPIhpi} & {\QTimeKFUhpi}\\
            $\text{int8} \to \text{int32}$  & $8 \times 1024 \times 8$ & {\QTimeKCYCbpi} & {\QTimeKPIbpi} & {\QTimeKFUbpi}\\
            \midrule
            $\text{fp32} \to \text{fp32}$   & $64\times 16 \times 64$ & {\QTimeSCYCspf} & {\QTimeSPIspf} & {\QTimeSFUspf}\\
            $\text{int32} \to \text{int32}$ & $64\times 16 \times 64$ & {\QTimeSCYCspi} & {\QTimeSPIspi} & {\QTimeSFUspi}\\
            $\text{int16} \to \text{int32}$ & $64\times 16 \times 64$ & {\QTimeSCYChpi} & {\QTimeSPIhpi} & {\QTimeSFUhpi}\\
            $\text{int8} \to \text{int32}$  & $64\times 16 \times 64$ & {\QTimeSCYCbpi} & {\QTimeSPIbpi} & {\QTimeSFUbpi}\\
          \bottomrule
        \end{tabular}
    {\centering \label{tab:q_perf}}
\end{table}
\begin{table}[t]
        \vspace{-1em}
        \centering
        \small
    \caption{Quadrilatero's area breakdown.}
          \begin{tabular}{lrr}
            \toprule
            Module & Area[$\mu m^2$] & \%        \\
            \midrule
            Controller       & \QCtrlArea & \QCtrlAreaP \\
            Register File    & \QRFArea   & \QRFAreaP   \\
            Permutation Unit & \QPUArea   & \QPUAreaP   \\
            Load-Store Unit  & \QLSUArea  & \QLSUAreaP  \\
            Systolic Array   & \QSAArea   & \QSAAreaP   \\
            \ \ $\vdash$ Combinational & (\QSAAreaComb) & \ \ (\QSAAreaCombP) \\
            \ \ $\vdash$ Sequential    & (\QSAAreaSeq)  & \ \ (\QSAAreaSeqP) \\
            \midrule
            Total            & \QTOTArea  & $100\%$   \\
          \bottomrule
        \end{tabular}
    {\centering \label{tab:q_area}}
\end{table}

We synthesize Quadrilatero in a 65-nm low-power technology node targeting the worst-case corner (SS, 1.08V, 125C). As shown in Table \ref{tab:q_area}, the {\QSAAreaCombP} of the area of Quadrilatero is dedicated to the combinational part of the \acrshort{MAC} units. 
The single-cycle latency \acrshort{FPU} limits the maximum frequency to 140 MHz.

We compare the post-synthesis \acrshort{PPA} (65nm, worst-case corner) of our architecture against Spatz and Spatz MX coupled with a scalar core and memory as described in Figure \ref{fig:main}. Vector processors achieve higher \acrshort{FPU} utilization as the \texttt{N} dimension increases, while Quadrilatero as the \texttt{K} dimension increases. So, for a fair comparison, we compare the execution of a $64 \times 64 \times 64 $ fp32 \acrshort{matmul}. We neglect the integer support of Quadrilatero in the comparison and use the same single-cycle latency \acrshort{FPU} module in all the architectures. 
Considering that Spatz supports more instructions than Quadrilatero, we only consider the \acrshort{PPA} of \acrshort{RF} and \acrshort{FPU}s in our comparison.

Configuring a vector processor with the same \acrshort{RF} bandwidth and number of \acrshort{FPU}s as Quadrilatero is not feasible. So, we carry out the following comparisons, where we configure Quadrilatero as described in Section 3, and consequently, it has 16 32-bit \acrshort{FPU}s, an 8$\times$4$\times$128-bit (4-Kibit) \acrshort{MRF}, and 4 32-bit memory ports:\\[0.1cm]
\textit{1) Quadrilatero vs. Spatz (same number of \acrshort{FPU}s})\\
Spatz has 16 32-bit \acrshort{FPU}s, a 32x512-bit (16-Kibit) \acrshort{VRF}, and 16 32-bit memory ports (Spatz has a higher \acrshort{RF} bandwidth and a larger \acrshort{RF}).\\[0.1cm]
\textit{2) Quadrilatero vs. Spatz (same \acrshort{RF} bandwidth)}\\
Spatz has 4 32-bit \acrshort{FPU}s, a 32$\times$128-bit (4-Kibit) \acrshort{VRF}, and 4 32-bit memory ports (Quadrilatero has 4$\times$ more \acrshort{FPU}s).\\[0.1cm]
\textit{3) Quadrilatero vs. Spatz MX}\\
Spatz MX has 4 32-bit \acrshort{FPU}s, a 32$\times$128-bit (4-Kibit) \acrshort{VRF}, 4 32-bit memory ports, and a 4x32 bits accumulator between the \acrshort{FPU}s and the \acrshort{VRF} to reduce \acrshort{RF} accesses.\\[0.1cm]
We show the results in Figure \ref{fig:comparison}.
Compared to \textit{1)}, \textit{2)}, and \textit{3)}, Quadrilatero improves the area efficiency, computed as the area-delay product (ADP), by {\AQsameFPUAEffP}, {\AQsameBWAEffP}, and {\AQMXAEffP}, respectively, due to the property of packing more \acrshort{FPU}s within the same \acrshort{RF} bandwidth without impacting the \acrshort{RF} size. Moreover, Quadrilatero is {\TQsameBW}$\times$ faster than \textit{1)} and \textit{2)} as it has $4\times$ more \acrshort{FPU}s  and {\TQsameFPU} slower than \textit{3)} while having just 25\% of its memory bandwidth and being {\AQsameFPU} smaller. Moreover, Quadrilatero saves {\EQsameFPU}, {\EQsameBW}, and {\EQMX} of energy (extracted at 100 MHz in the typical corner\textemdash TT, 1.20V, 25C) compared to \textit{1)}, \textit{2)}, and \textit{3)}, respectively, due to the reduced number of \acrshort{RF} accesses. 
\section{Conclusions}
In this paper, we analyzed the advantages of a matrix ISA over a vector ISA, showing how it can alleviate \acrshort{VRF} bandwidth requirements and reduce costly register-file accesses during \acrshort{matmul}s.
We designed Quadrilatero, an open-source area-efficient RISC-V programmable matrix coprocessor for low-power edge applications, to exploit these advantages for a more efficient AI computation at the edge. We evaluated its post-synthesis \acrshort{PPA} in a 65-nm technology node, showing that it requires only {\QTOTAreaMM} $mm^2$, of which {\QSAAreaCombP} are employed by the combinational logic of the \acrshort{MAC} units, that it can reach up to {\QTimeKFUspf} of \acrshort{FPU} utilization. Compared to a state-of-the-art vector processor and a hybrid vector-matrix processor, it improves the area efficiency at 140 MHz up to {\AQMXAEff}$\times$ and the energy consumption at 100 MHz up to {\AQsameFPUAEffP}.

\begin{acks}
This work is funded in part by the dAIEDGE project supported by the EU Horizon Europe research and innovation program under Grant Agreement Number: 101120726.\\
We thank Mr. Julien François De Castelnau for the compiler support.
\end{acks}

\bibliographystyle{ACM-Reference-Format}
\bibliography{main}
\end{document}